\documentclass{article} 
\usepackage{graphicx}
\title{Underpinning the universe: its scales, holography and fractality
}

\author{Antonio Alfonso-Faus\thanks{E-mail: aalfonsofaus@yahoo.es} \\ 
Escuela de Ingenier\'{\i}a Aeron\'autica y del Espacio \\
Plaza del Cardenal Cisneros, 3, 28040 Madrid, Spain\\
\and M\`arius Josep Fullana i Alfonso\thanks{E-mail: mfullana@mat.upv.es} \\
Institut de Matem\`atica Multidisciplin\`aria, \\
Universitat Polit\`ecnica de Val\`encia \\
Cam\'{\i} de Vera s/n, 46022 Val\`encia, Spain}

\begin{document}

\maketitle

Pacs: 98.80.-k; 04.70.Bw; 03.65.-w


\abstract{
We expand on the general concept of a universe. We identify physics as a unit applied to a universe. Then we generalize the concept of a quantum black hole, and apply it to the unit of a universe. We find that only one parameter, the Pin, is needed to define all its physical properties. Here we present three significant quantum black holes, three scales: Planck's, sub-Planck and our own universe as a whole. Then we revise the holographic and fractal properties, and propose a sequential growing process to explain the evolution and the basic structure of our universe.}

\section{Introduction}

The concept of a universe is a very intuitive one: it contains everything that is known (and not yet known too) to exist. The field of physics is known as the science of matter (energy). A universe is then the field of physics by excellence, and as a science it has its own name: cosmology. The idealization of a universe is contained in the Cosmological Principle, and it is of the outmost importance. Today we deal with three components of our universe: what we see and touch (4\%), dark matter (24\%) and dark energy (72\%) both of them neither seen nor touched .The last two constituents are named as "dark" because we do not know what they are, what they are made of. Then we can say that most of our universe is unknown. The scientific literature in cosmology is very rich, abundant and ingenious. We present here our contribution based upon the idea that, not knowing very well what we are talking about, we may try to go on idealizing and, perhaps, using a bit more the concepts introduced by mathematics. We introduce the concept of the Pin of a quantum black hole, a universe in itself. Only one parameter is needed to completely define it. This may simplify things. A universe seems to be holographic and fractal and its underpinning may be that it is a sequential growing process applied to the Planck´s quantum black hole. An analysis of the concepts discussed in our article from the statistical point of view is intended in the near future \cite{VVU}.

\section{Unification of physics}
The Universe is the unity for excellence; all its properties are integrated, entangled. It has structure. And its study is based, very successfully, in a principle due to Milne and coined by Einstein: The Cosmological Principle. Its content is simple and "democratic": 1) any place in the universe is equivalent to any other, they cannot be distinguished at the same instant of time. 2) If we observe the universe from one place in all directions, at the same instant of time, we will always see the same structure in any direction. These two ideas are explained by saying, equivalently, that the universe is homogeneous and isotropic. They can be taken as a simplification of "reality", a simplification that works. But it is also possible to think that they are a correct idealization of reality. If we observe the world around us we do not exactly see this. If we move, things that appear are different to us, and if we look from a place in different directions, we also see everything different. The point is that the Cosmological Principle refers to long distances, looking far away and considering averages which do not take into consideration particularly close details. We can see this principle as an equivalent to the following thought: in an ocean full of islands where each one is different from the other ones and are randomly distributed, they all have in common that they are united underneath, by the land, by the depth of the ocean which is their base. This common property for all the islands is equivalent to the Cosmological Principle, valid for the entire universe. And it is the same as to say that the universe has no centre. Or that any place in the universe can be considered to be at its centre. We could say that we have infinite centres. 

Nevertheless we study the behaviour and properties of our universe under the prism of what we call General Physics, divided in various self contained and separated "branches". A reductionist approach, but it gives good results as well. The majority of these branches are only valid under a reduced field of application. Our mind has divided the unity of the cosmos in many particular branches of knowledge, included all of them in General Physics. We can enumerate them as follows:

(a) Classical Mechanics (Newton, $G$)
(b) Relativity (Einstein, $G$ and $c$)
(c) Quantum Mechanics (Schr\"odinger, $\hbar$)
(d) Electromagnetism (Maxwell, $e$)
and (e) Thermodynamics (Boltzmann, $k$).

Here we have indicated the characteristic constant for each branch: $G$, the universal constant of gravity, $c$, the speed of light, $\hbar$, Planck's constant, $e$, the electric charge of the electron and $k$, Boltzmann's constant. Nowadays we have indications of the possibility to achieve the confluence of all these branches into one point/object only: the quantum black hole. This object fuses in a single unit the following two properties:

A)	It is a black hole

B)	It is a quantum

And we will see that this object is a universe in itself.

A)	The black hole

The concept of a black hole, being already well known, comes from considering a mass m inside a space of size $r$ (for example a sphere of radius $r$). The trick to get a black hole of mass m and size r consists of having the properties, mass and size, related in the following way:

\begin{equation}
\label{eq.1}
2Gm/c^2 = r 
\end{equation}

This is the relation that is known as the Schwarzschild radius. If we define a system of units such that $2G = c = 1$, which is always possible, then eq. (\ref{eq.1}) would simply be

\begin{equation}
\label{eq.2}
m = r 
\end{equation}

And if we consider the characteristic time t of this black hole, being $r = ct$, we would have 

\begin{equation}
\label{eq.3}
m = r  = t
\end{equation}

In conclusion: The physical properties of a black hole, its mass, length and characteristic time, can be considered as equivalent, 
and therefore it is possible to define the black hole by only one of these  by taking $2G = c = 1$.

This conclusion can be clearer considering the dimensions of $G$ and $c$:

\begin{equation}
\begin{array}{c}
[G] = \hbox{length}^3/(\hbox{mass}  \times  \hbox{time}^2) \\
\hbox{and} \ \ \ [c] = \hbox{length} / \hbox{time} 
\end{array}
\label{eq.4}
\end{equation}

Here we see that when we make $c = 1$ we are matching the dimensions of length and time. And when we make $2G = 1$, eliminating already length = time, we get that length is also equivalent to mass. The first scientist to develop the concept of black hole was the English geologist John Michell. He sent a letter with this idea to the Royal Society in 1783. We should give the name of Michell to the three properties of a black hole, length = mass = time. Nevertheless we choose the name of this unit as the {\bf Pin}, instead of Michell, to follow a physical/mathematical context. So the $Pin$ unit is the number that corresponds to the length, mass and time of a black hole, when we consider $2G = c = 1$. A $Pin$ unit refers clearly to a black hole defined by only one property: its size, or its mass, or its characteristic time. It is its personal identification number.

We can include now, following this denomination of the $Pin$ unit, the angular momentum of the black hole, if it has any. The dimensions of the angular momentum are:

\begin{equation}
\label{eq.5}
\hbox{mass} \times \hbox{speed}  \times  \hbox{length} = [Pin]^2
\end{equation}

We can include the electric charge of the black hole, if there is any. Since the fine structure constant is dimensionless we have $hc/e^2 = [Pin]^0$, and having the Planck's constant h the dimensions of the angular momentum, as in eq. (\ref{eq.5}), the electric charge e has the dimensions of the $Pin$ unit: 
          
\begin{equation}
\label{eq.6}
[hc] = [e^2]    ,  \ [\hbar] = [Pin]^2 \Rightarrow  [e] = Pin   
\end{equation}        

So the dimensions of the black hole are defined only by one property: length, mass, time, electric charge or angular momentum. Thus, a black hole is completely defined by only one of its properties, its $Pin$, its personal identification number.

B)	The quantum black hole

Let us analyze now what is meant by a quantum property in general. At the base of this concept is the Compton wave length of an object, defined by the mass $m$ as

\begin{equation}
\label{eq.7}
\lambda_c = \hbar / m c
\end{equation}

We have already defined the radius of Schwarzschild in eq. (\ref{eq.1}) and we can now force this length to be of a quantum nature, as defined here, equal to the Compton wavelength. Then combining eqs. (\ref{eq.1}) and (\ref{eq.7}) we have 

\begin{equation}
\label{eq.8}
2Gm/c^2 = r = \hbar / m c = ct
\end{equation}

The choice $2G = c = 1$ applied to eq. (\ref{eq.8}) gives 

\begin{equation}
\label{eq.9}
m = r = h/m = t 
\end{equation}

An {\it object} having these characteristics is defined as a quantum black hole. The quantum property is the one that introduces the constant $\hbar$ in the definition. We then have

\begin{equation}
\label{eq.10}
\hbar = m^2 = r^2 = t^2 == [Pin]^2
\end{equation}   

In eq. (\ref{eq.7}), if we consider the mass m to correspond to a quantum of gravity \cite{AAF1}, i.e. the minimum mass known to have a physical meaning, we obtain a quantum of size equal to the size of the universe today. If we go to the other extreme, if we consider that the mass m is equal to the mass M of the visible universe (the biggest known), then the quantum has the size of the gravitational radius of the quantum of gravity, the smallest size with a physical meaning. We are clearly jumping from a universe to the other, from one black hole to the other, and besides ¡ they are both of a quantum nature in a general sense! The solution to this paradox is possible to carry out by generalizing the "constant" of Planck h. We know that this constant has dimensions of $[Pin]^2$ as we saw in eqs. (\ref{eq.5}) and  (\ref{eq.10}). To define any quantum black hole we have enough with a value of its Pin. And this also allows us to define ITS GENERALIZED PLANCK'S CONSTANT AS $(PIN)^2$.

C)	Planck's quantum black hole and its $Pin \ h^{1/2}$  

Since a quantum black hole is completely defined by one parameter, its $PIN$, we can define quantum black holes from this property. There are three that are very important. First we have the Planck's quantum black hole, with its Pin defined by a size of $\sim  10^{-33} cm$ (or equivalently by a mass of $ \sim 10^{-5} g$). It has a mass, length and time that are obtained from the most simple combination of the three constants $G$, $c$ and $\hbar$. Of course these three properties come from the conditions:
$2Gm_p/c^2 = l_p = ct_p = \hbar / (m_p c)$, i.e.
\begin{equation}
\label{eq.11}
\begin{array}{c} 
                           l_p = (G\hbar/c^3)^{1/2} \approx \hbar^{1/2}  \approx 10^{-33} cm \\
                            m_p = (\hbar c/G)^{1/2} \approx  \hbar^{1/2} \approx  10^{-5} g           \\                            
                           t_p = (G\hbar/c^5)^{1/2} \approx \hbar^{1/2}  \approx 5 \times10^{-44} s
\end{array}
\end{equation}

\noindent
$12 \times (\hbox{electric charge} e)_p \approx  (\hbar c)^{1/2} = \hbar^{1/2}$  (The $Pin$ of Planck).

The factor 12 is approximately $137^{1/2}$ and is directly related to the fine structure constant  $\alpha = e^2/\hbar c \approx  1/137$. Normally the system of units eq. (\ref{eq.11}) is used as the so called natural system. We see that to take this system as a reference unit is exactly equal as to take the Planck's constant, its $Pin^2$, equal to 1.

D)	The sub-Planck quantum black hole and its $Pin \ 10^{-61}$ 

We define the quantum black hole called sub-Planck as the one obtained from Planck's eq. (\ref{eq.11})  reducing its Pin by the scale factor $10^{61}$. Doing so, we obtain a quantum black hole with a very interesting physical meaning: as we already said it is the suitable candidate to be the quantum of gravity \cite{AAF1}. We can see now what would happen if we take as the reference unit this quantum black hole. The units would now be:
\begin{equation}
\label{eq.12}
\begin{array}{c} 
                           l_{sp} = 10^{-61} (G\hbar/c^3)^{1/2} \approx 10^{-61} \hbar^{1/2}  \approx 10^{-94} cm \\
                            m_{sp} = 10^{-61} (\hbar c/G)^{1/2} \approx  10^{-61} \hbar^{1/2} \approx  10^{-66} g           \\                            
                           t_{sp} = 10^{-61} (G\hbar/c^5)^{1/2} \approx 10^{-61} \hbar^{1/2}  \approx 5 \times10^{-105} s
\end{array}
\end{equation}
\noindent
$12 \times (\hbox{electric charge})_{sp} \approx  10^{-61} (\hbar c)^{1/2} = 10^{-61} \hbar^{1/2}$  (The $Pin$ of sub-Planck).
We see that if we take the $Pin$ of Planck as a unit ($\hbar = 1$) the $Pin$ of sub-Planck is $10^{-61}$.

E)	A very special quantum black hole : the universe and its $Pin 10^{61}$

Finally we define the quantum black hole, which we call our universe \cite{AAF2}, as the one obtained from the Planck's one (\ref{eq.11})  multiplying its $Pin$ by the scale factor $10^{61}$. This is equivalent to obtain this quantum black hole from the sub-Planck one (\ref{eq.12})  multiplying its $Pin$ by the scale factor $10^{61}$ twice, that is to say, by the factor $10^{122}$. Doing so, we obtain a quantum black hole with a much more interesting physical meaning: it is our visible universe. We can see now what would happen if we take as the unit of reference this quantum black hole. The units would now be:
\begin{equation}
\label{eq.13}
\begin{array}{c} 
                           l_{u} = 10^{61} (G \hbar/c^3)^{1/2} \approx 10^{61} \hbar^{1/2}  \approx 10^{28} cm \\
                            m_{u} = 10^{61} ( \hbar c/G)^{1/2} \approx  10^{61} \hbar^{1/2} \approx  10^{56} g           \\                            
                           t_{u} = 10^{61} (G \hbar/c^5)^{1/2} \approx 10^{61} \hbar^{1/2}  \approx 5 \times 10^{17} s
\end{array}
\end{equation}
\noindent
$12 \times (\hbox{electric charge})_{u} \approx  10^{61} (\hbar c)^{1/2} = 10^{61} \hbar^{1/2}$  (The $Pin$ of the universe).
We can see that taking the Planck's quantum black hole as the unit ($\hbar = 1$) is equivalent as to say that the $Pin$ of the universe equals $10^{61}$. 

If we combine the relations of Weinberg and Zel'dovich \cite{AAF3} we obtain the result that the cosmological constant $\Lambda$, due to Einstein, is of the order of the inverse squared size of the universe. Then its dimension is $[Pin]^{-2}$. Here we have solved the annoying problem of this constant, which has a discrepancy in its value of a factor of the order of $10^{122}$ when comparing cosmology and the standard theory of particles. This discrepancy is completely fictitious since it is due to a change of scale, and this gives us the square of $10^{61}$, which is equal to the value $10^{122}$, the square of the relation of Pins, see \cite{AAF3}.

We point out here that we have found a very promising relation for {\it connecting}, general relativity with quantum mechanics: From one side we have the generalized constant of Planck. From the other side we can generalize the cosmological constant $\Lambda$ too. Combining the relations of Weinberg and Zel'dovich \cite{AAF3} we obtain the result of $\Lambda \approx  1/R^2$. That is to say that $ [\Lambda] = [Pin]^{-2}$. The product of the two generalized constants, $\hbar$ and $\Lambda$, is of the order of 1. Therefore this product has the dimension $[Pin]^0$. It is therefore an invariant scalar, as $G$ and $c$. Then we have now the possibility of introducing the constant of Planck in the cosmological equations of Einstein: $\hbar$ is of the order of the inverse of $\Lambda$.

For these three quantum black holes the value of their $Pins$ runs from the unit (if we take it arbitrarily for the Planck's quantum black hole), to the value of $10^{-61}$, for the sub-Planck, up to the third one, our universe, multiplying twice the sub-Planck or once Planck's by $10^{61}$.

We have said that the quantum property of a system of mass $m$ (be it or not a black hole) is obtained by introducing the Planck's constant h by means of the definition of the Compton wavelength. In a quantum sense, this wavelength can be interpreted as the quantum size of the mass m and it is defined as we already know by eq.  (\ref{eq.7}).

The fact is that we can oblige this quantum size to be exactly equal to the size $r$ of the black hole of mass $m$. By doing so, and generalizing the constant of Planck $\hbar$ with the denomination $\bar{H}$, the universal relations are obtained which define any quantum black hole:

\begin{equation}
\label{eq.14}
2Gm/c^2 = r = \bar{H} / m c = [Pin]^1
\end{equation}  

With $2G = c = 1 = [Pin]^0$  we obtain $m = r = \bar{H} / m  = [Pin]^1$,
that is to say       

\begin{equation}
\label{eq.15}
\bar{H} = m^2  = r^2 = [Pin]^2
\end{equation}  

And we have also the relations 
\begin{equation}
\label{eq.16}
\bar{H} \Lambda = [Pin]^0  \hbox{ (order unity)  and }  \bar{H} = 1 /  \Lambda  = [Pin]^2 
\end{equation}
                               
If we consider now the thermodynamic aspect, we can introduce a temperature known as Planck's temperature by dividing the energy of the Planck's quantum black hole by Boltzmann's constant $k$. This temperature turns out to be approximately of  $\sim 10^{32}  K$. It can also be taken as a reference unit, considering Boltzmann's constant $k$ equal to 1, as $G$, $c$ and $\hbar$ for this Planck's object. As we have previously seen, the product of the constant of Planck $\hbar$ by the light speed $c$ is dimensionally equal to the square of an electric charge. This electric charge of the quantum black hole is of the order of 12 times the electric charge of the electron $e$. With all this in mind, we obtain a system of units defined by:

\begin{equation}
\label{eq.17}
2G = c = \hbar = k = 1 \approx  12e
\end{equation}

By making all these constants equal to one we obtain the scale of Planck. Now we can consider that the five branches of physics, that our mind has considered separately, are all integrated in only one object: The quantum black hole of Planck. But this also happens whatever quantum black hole you consider, whatever its $Pin$.

Following the exposition in \cite{AFIS}, the dimension of Boltzmann's constant $k$ is equal to the one of Planck's $\hbar$ i.e. $[Pin]^2$. Therefore the entropy $S$, which has the dimensions of  $k$ or $\hbar$, that is an energy (energy is $mc^2$ and therefore has $Pin$ 1 dimension) divided by a temperature $T$ 

\begin{equation}
\label{eq.18}
\begin{array}{c}
[S] = [ \hbox{Energy} ] / [T] = [Pin] / [T] = [Pin]^{2} \\
\ [T] = [Pin] / [Pin]^{2} = 1 / [Pin]
\end{array}
\end{equation}

This expression is very important from the cosmological point of view. It means that the product of temperature by time ($Pin$ 1) is dimensionless. Therefore the expansion of the universe, or its age (both with $Pin$ 1) imply a cooling effect, a decrease in temperature. And surprisingly, having the temperature a statistical context, so does time.

The existence, be it real or virtual, of the Planck's quantum black holes has been proposed to be a characteristic of the fluctuations of the quantum vacuum: the supposed "foam" as the structure of such a vacuum which corresponds to the possible fluctuations of space-time, at this scale. If we use this idea, simplifying it to only one fluctuation, we can propose that the universe was shaped out of an initial seed, the quantum black hole of Planck, which expanded (or replicated) up to the size that the universe has today. We imply this size to be its visible size that goes on expanding and in an accelerated way as of today. If the universe was born from a quantum black hole of Planck, an initial fluctuation, which expanded up to the present, it is reasonable to think that it has always preserved the property of being a quantum black hole. We have initiated this idea implicitly by stating that "The universe is one, all its properties are integrated, entangled". Of course the quantum black hole of Planck is also one, and in it all physical properties go together. It is in a sense also a universe. The same happens to our universe, everything in it is integrated. It is plausible then to think that the expansion (or replication) of the first fluctuation has led it to its present state. A proof in favour of the acceptability of this idea is the cosmological scale factor: this dimensionless factor, which is of the order of $10^{61}$ today, applied to all the physical properties of the quantum black hole of Planck gives us all the physical properties of our universe nowadays. For example, the values of length, mass and time in eq. (\ref{eq.11}), multiplied by this scale factor give us the characteristics of our visible universe nowadays: size 
$\sim 10^{28} cm$ , mass $\sim 10^{56} g$, and age $\sim 4 \times 10^{17} s$. It does not seem probable that this result is just a coincidence.

\section{The holographic principle, the information and the bit}

The first two scientists who advanced the holographic principle were 't Hooft \cite{GH} and Susskind \cite{LS}. In short, this principle establishes the following property: given a volume 
$V \approx  R^3$, of  linear size $R$, and confined by a surface  $A \approx  R^2$, all the information contained inside the volume $V$ (evidently of  dimension 3) is  equivalent to the information that can be defined and analyzed in the surface $A$ (dimension 2) which encloses $V$. We have a practical example of this principle in the holographic photography: it has two dimensions, and it can be converted into a three dimensional view. In this way, the information contained in $A$ is equivalent to the one in a space of three dimensions, as we observe in our universe.

Considering the universe as a black hole, its entropy $S$ calculated by Hawking \cite{SWH} for a black hole is given by the expression

\begin{equation}
\label{eq.19}
S = 4 \pi k / \hbar c \ G M^2
\end{equation}

Since $k$ and $\hbar$ have the same $Pin$ dimension \cite{AFIS}, and $c$ and $G$ have the zero dimension, the dimension of the entropy $S$ is of $ [Pin]^2$, as we have already seen.  In Planck's units this entropy has the value
\begin{equation}
\label{eq.20}
S \approx (M/m_p)^2 \approx  10^{122}
\end{equation}
\noindent
where $M \approx  10^{56} g$ and $m_p \approx  10^{-5}g$. Having used in this definition the total mass of the universe $M$ it is clear that we are considering a finite total mass inside a volume $V \approx  R^3$ (dimension 3). If we have in mind that our universe is a black hole, we have (considering that mass and size have the same dimension, $[Pin]$) using eq. (\ref{eq.19})
\begin{equation}
\label{eq.21}
S \approx k (R/l_p)^2 \approx  10^{122}
\end{equation}
\noindent
where $R \approx  10^{28}cm$ and $l_p \approx  l0^{-33}cm$. Clearly the consideration of the universe as a black hole leads to the validity of eq. (\ref{eq.19}). 
The numerical equality of eq. (\ref{eq.20}) and eq. (\ref{eq.21}) implies that the universe is holographic. According to the definition of bit \cite{AAF1}, 
identified as having a mass $m_g = \hbar /cR  \approx  10^{-66} g$ (the mass at the sub-Planck's scale), the number of bits $N$ contained in the universe is 
\begin{equation}
\label{eq.22}
N \approx M/m_g \approx  10^{122}
\end{equation}

The equality of the three big numbers in eqs. eq. (\ref{eq.20}), (\ref{eq.21}) and (\ref{eq.22}) is completely in line with the consideration that the universe is a quantum black hole, holographic, and that it contains an information equivalent to its entropy, a value of $10^{122}$ bits. 
                  
\section{Fractality of the universe}

A definition of fractality for a set of systems of mass $M_i$  and size $r_i \ ( i = 1,2,\dots) $ has the expression
\begin{equation}
\label{eq.23}
M_i/ (r_i )^d =  \hbox{constant} 
\end{equation}
\noindent
where $d$ is the dimension of fractality. Any black hole has its mass proportional to its size \cite{AAF1}. Then all the black holes are fractal, with dimension $d = 1$. 
Besides they are holographic because their entropy, given by Hawking, eq. (\ref{eq.19}), is proportional to the square of their mass, i.e., the square of their size. 
Then the quantum black holes that we have introduced here, the universe, the Planck's one and the sub-Planck one, are also holographic and fractal.

\section{Short history of cosmology as a science}
The two cosmological equations of Einstein were obtained nearly 100 years ago. With them began the scientific era of cosmology. The theory of general relativity of Einstein continues to be successfully validated day by day. And with this we also have the validation of his two cosmological equations. From the static model of the universe of Einstein very soon we changed to a universe in expansion, thanks to the works of Hubble. And from about 10 years ago we also have the conviction 
that the universe is not only expanding but it is doing it in an accelerated way  \cite{SPG} and  \cite{OATD}. 
An initial inflation, an exponential expansion, is something that it is still being discussed today. From another side we have the apparent necessity of postulating the existence of dark matter, in order to explain certain kinematic behaviour of some stars and galaxies. And we also have the necessity of postulating the existence of dark energy, whose pressure not only cancels the attraction of gravity but even more, it expands the universe.
The cosmological parameters, which are measured by the observations from different kind of telescopes, do not satisfy the first cosmological equation of Einstein. This is very embarrassing. It is necessary to add something more, something which must be out there and which has not yet been considered. Of course this {\it something} is responsible, to a certain extent, of the accelerated expansion that is telling us that gravity is being overcome. And it may be responsible for a relatively fast end of the universe. We will see it in the next section.

\section{The {\it magic} equation and its integration}

Since with present observations it is not possible to satisfy both equations of Einstein at the same time, we have proceeded to approach the cosmological problem from a purely kinematic point of view. We are not going to need the equations of Einstein. And we are going to use only the value of two parameters: the constant of Hubble $H$, which is already known with enough precision, and the deceleration parameter $q$ (which today is negative, acceleration). The kinematic definition of these two parameters is:

\begin{equation}
\label{eq.24}
\begin{array}{c}
H(t) = R'/R \\
q(t) = - R'' R / (R')^2
\end{array}
\end{equation}

Here the derivatives are relative to time t. Combining these two mathematical equations we obtain the next one
\begin{equation}
\label{eq.25}
 H' + (1+q)H^2 = 0
\end{equation}

\begin{figure}
\includegraphics[]{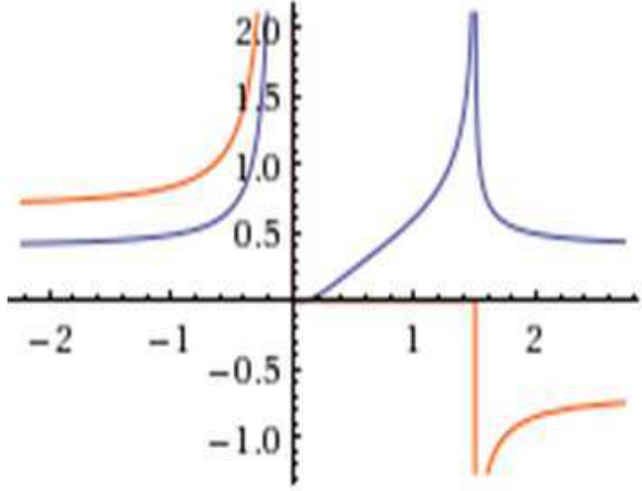}
\caption{Plot of the expansion scale factor, $R$, as a function of $x=t/t_0$. 
The blue/red line correspond to the real/imaginary part of eq. (\ref{eq.26}) integration.}
\label{fig.1}
\end{figure}

We call this equation {\it magic} because its integration gives us all the kinematic parameters of the universe: 
The scale factor of expansion $R(t)$, the speed of expansion $R'(t)$ and the acceleration of the expansion $R''(t)$. 
The first integration of eq. (\ref{eq.25}) can be obtained by expressing it in the following more convenient way
\begin{equation}
\label{eq.26}
d (1/H)/d t  = 1 + q(t)  
\end{equation}
                                
The function $q(t)$ is experimentally well known. The integration of eq. (\ref{eq.26}) is immediate, 
with the constant of integration determined by the present value of the product $(Ht)_0 \approx  1$.  
A second integration gives us the function $R(t)$ with the constant of integration as determined by the present value of $q_0 \approx  - 0.67$ . 
Taking the axis of abscissa as the parameter $x = t/t_0$ , relative to the age of the universe today $t_0 \approx  1.37  10^{10}$ years, and in the ordinate axis the scale factor referred to is present value $R(t_0)  \approx  10^{28} cm$, we obtain the graphic of fig. \ref{fig.1}.

We see in this graphic that the expansion of the universe is almost linear until now, 
with a small acceleration, and with an end of the universe for $x \approx  1.5$ where we have an expansion to infinite, disintegration.
  
\section{The parallelism between Human Being and its Universe}

Sometime in the past we have commented that Human Being seems to have been created resembling his Universe, in some way. And every day one finds more traces pointing towards this conclusion. The abundance of water in the universe is so high that it could be alike to the percentage of water in our bodies, an idea to be checked in the future. The structure of the neurons in our brain is much more alike to the filamentous structures of the universe observed by the telescopes. And now we see that if function $q(t)$ is confirmed as it is known today, the universe has a limited life: it has a birth and it will disappear at the end of about $2 \times 10^{10}$ years.  It had a beginning, it developed in a wonderful way and perhaps it will disintegrate in a finite time as Human Being does. Some works indicating that the universe may have a limited life are \cite{BMcI,RRC,CKW,SP,ZiL}.

\section{The process of a sequential growing} 

We have seen that the Pin that we have defined for the quantum black holes defines the dimensions: length, mass, time, electric charge, and so on. There are three additional physical properties which have a $Pin$ as $[Pin]^2$: The entropy, the angular momentum and the inverse of the cosmological constant $\Lambda$. On the other hand, we have the universal constants which must be unit in this scale $[Pin]^0, \ G, \ c, \ \hbar /k,$ the products $\bar{H} \Lambda$ and $S\Lambda$. The universal constants remain constant, dimensionless, of the order of unity in this scale. The physical properties that are linear with its $Pin$ continue to be so in the process of sequential growing, at quantum jumps \cite{SPG}. The physical properties which have their $Pin$ squared form a chain in a sequential growing in two dimensions: horizontal and vertical, each one being linear \cite{SPG}. The case for the entropy is clear, having in mind the holographic principle: it has a sequential growing in two dimensions, proportional to the area which limits the volume under consideration. In particular the chain of elements formed by a casual set of quantum black holes, like Planck's, grows in two dimensions equally, horizontal and vertical. Each one identifies our universe, with a total of $10^{61}$ elements (the scale of the universe, the one that turns the Planck's quantum black hole in our universe). For the readers who want to go deeper in the mathematical meaning of these chains there is a very recent work \cite{SPG} which contains many references really interesting. These chains explain the evolution and the development of our universe.

\section{Conclusions}
It is possible to define a universe with its physics unified, and to identify it with a quantum black hole. Following this line of thought our universe has a three level of quantum black holes: the universe itself, Planck's and sub-Planck quantum black holes. While the Planck's one can explain our present universe, by a process of sequential growing, the sub-Planckian level may be identified with a lattice structure of the "vacuum". Each cell would have a Planck's size with a sub-Planck quantum black hole, a bit, at its centre.

\section{Ackwnoledgements}
There are three acknowledgements we must express: to Da. Maria Cecilia Est\'efan Arbel\'aez for her suggestion of denominating as Pin all the properties of the quantum black hole. To Dr. Stanley P. Gudder for his confirmation of the possibility of chains in the process of sequential growth, specifically referring to quantum black holes of Planck. And last, our acknowledgment to the organizers of the mathematical programs in internet, the Wolfram Mathematica Online Integrator, which we have used again and again for the integration and graphics of the {\it magic} equation.

\end{document}